# Growth of the Moon Due To Bodies Ejected from the Earth

S. I. Ipatov*

*Vernadsky Institute of Geochemistry and Analytical Chemistry, Russian Academy of Sciences, Moscow, Russia*
*\*e-mail: siipatov@hotmail.com*

**Abstract**—The evolution of the orbits of bodies ejected from the Earth has been studied at the stage of its accumulation and early evolution after impacts of large planetesimals. In the considered variants of calculations of the motion of bodies ejected from the Earth, most of the bodies left the Hill sphere of the Earth and moved in heliocentric orbits. Their dynamical lifetime reached several hundred million years. At higher ejection velocities $v_{ej}$ the probabilities of collisions of bodies with the Earth and Moon were generally lower. Over the entire considered time interval at the ejection velocity $v_{ej}$, equal to 11.5, 12 and 14 km/s, the values of the probability of a collision of a body with the Earth were approximately 0.3, 0.2 and 0.15–0.2, respectively. At ejection velocities $v_{ej} \leq 11.25$ km/s, i.e., slightly exceeding a parabolic velocity, most of the ejected bodies fell back to the Earth. The probability of a collision of a body ejected from the Earth with the Moon moving in its present orbit was approximately 15–35 times less than that with the Earth at $v_{ej} \geq 11.5$ km/s. The probability of a collision of such bodies with the Moon was mainly about 0.004–0.008 at ejection velocities of at least 14 km/s and about 0.006–0.01 at $v_{ej} = 12$ km/s. It was larger at lower ejection velocities and was in the range of 0.01–0.02 at $v_{ej} = 11.3$ km/s. The Moon may contain material ejected from the Earth during the accumulation of the Earth and during the late heavy bombardment. At the same time, as obtained in our calculations, the bodies ejected from the Earth and falling on the Moon embryo would not be enough for the Moon to grow to its present mass from a small embryo moving along the present orbit of the Moon. This result argues in favor of the formation of a lunar embryo and its further growth to most of the present mass of the Moon near the Earth. It seems more likely to us that the initial embryo of the Moon with a mass of no more than 0.1 of the mass of the Moon was formed simultaneously with the embryo of the Earth from a common rarefied condensation. For more efficient growth of the Moon embryo, it is desirable that during some collisions of impactor bodies with the Earth, the ejected bodies do not simply fly out of the crater, but some of the matter goes into orbits around the Earth, as in the multi-impact model. The average velocity of collisions of ejected bodies with the Earth is greater at a greater ejection velocity. The values of these collision velocities were about 13, 14–15, 14–16, 14–20, 14–25 km/s with ejection velocities equal to 11.3, 11.5, 12, 14 and 16.4 km/s, respectively. The velocities of collisions of bodies with the Moon were also higher at high ejection velocities and were mainly in the range of 7–8, 10–12, 10–16 and 11–20 km/s at $v_{ej}$, equal to 11.3, 12, 14 and 16.4 km/s, respectively.

**DOI:** 10.1134/S0038094624010040

## INTRODUCTION

During the formation of the Earth and the Moon, various planetesimals (bodies) fell on them. When such bodies collided with the Earth, matter could be ejected from it, some of which fell on the Earth and the Moon. Armstrong et al. (2002) proposed studying lunar rocks to find material ejected from the Earth during the late heavy bombardment and to better understand the history of the Earth. It is believed (Bottke and Norman, 2017) that the late heavy bombardment occurred approximately 4.2–3.5 billion years ago and could have been caused by the end of the process of planet accumulation and the migration of the giant planets. Recently, falls of large bodies capable of causing the ejection of material from the Earth beyond its Hill sphere are rare. The work of (Gattacceca et al., 2023) claims that the meteorite 13188 found in Northwest Africa is a terrestrial meteorite that fell back to the Earth. This paper discusses the probability of bodies ejected from the Earth falling onto the Moon, the velocity of collisions of ejected bodies with the Earth and the Moon, as well as the growth of the Moon embryo when bodies ejected from the Earth fall on it. Below, we first provide an overview of some papers on the topic of the paper.

### Moon Formation Models

There are different models for the formation of the Moon. We provide reviews of these models in (Ipatov, 2018), in section 2.1, in (Marov et al., 2019), and in section 3.2 of a large review (Marov and Ipatov, 2023).

The giant-impact theory is very popular (for example, Hartmann and Davis, 1975; Cameron and Ward, 1976; Canup and Asphaug, 2001; Canup, 2004, 2012; Canup et al., 2013, 2021; Cuk and Stewart, 2012; Cuk et al., 2016; Barr, 2016). In this model, it is believed that the Moon was formed by the ejection of the Earth's silicate mantle when the Earth collided with a body of the size of Mars. Numerous variants of collisions with different impactor masses and collision velocities were considered. The giant-impact model faces certain difficulties, primarily of a geochemical nature. In particular, criticism of this model was given in numerous papers of Galimov and his coauthors (Galimov et al., 2005; Galimov, 1995, 2008, 2011, 2013; Galimov and Krivtsov, 2012; Vasiliev et al., 2011). For example, Galimov (2011) believed that the giant- impact theory cannot explain the fact that a shift in the isotopic composition between lunar and terrestrial matter was not detected, since the material ejected during the giant-impact should consist of 80–90% vapor, and when the melt evaporates, the isotopic compositions of K, Mg, and Si can vary noticeably.

In the multi-impact model (Ringwood, 1989; Vityazev and Pechernikova, 1996; Gorkavyi, 2004; 2007; Svetsov et al., 2012; Citron et al., 2014; Rufu and Aharonson, 2015; Rufu et al., 2017, 2021) considered the formation and growth of the Moon embryo mainly due to the material of the Earth's crust ejected from the Earth embryo during its several collisions with the bodies of the protoplanetary disk. Rufu et al. (2017) considered that there were about 20 impactors. Their calculations considered impactor masses from 1 to 9% of the Earth's mass and collision velocities from $v_{esc}$ up to $4v_{esc}$, where $v_{esc}$ is the escape velocity from the Earth's surface. The algorithm used in their calculations performed hydrodynamic modeling. In various versions of the megaimpact and multi-impact models, the ejected matter (partially in the form of vapor) formed a disk from which the Moon embryo was formed. In the multi-impact model, after each collision, a new body was formed from the disk, which, moving away from the Earth, collided and merged with the previous embryo of the Moon. Rufu et al. (2017) noted that the lunar embryo formed at a distance greater than the Roche limit (three Earth radii).

The coaccretion theory considers the formation of the Moon from a swarm of small bodies. A monograph and a number of papers by Ruskol (1960, 1963, 1971, 1975) are devoted to this model. The main source of the near-Earth swarm of bodies in this model is the capture of particles of the preplanetary disk during their collisions ("free with free" and "free with bound"). In (Afanasyev and Pechernikova, 2022) it was found that as a result of pair collisions of preplanetary bodies, a near-Earth swarm with a mass of about $10^{-5}$ the mass of the present Moon could be formed. The authors suggested that this swarm could serve as a trigger for further accretion due to collisions of bodies of the protolunar swarm with bodies from the planet's feeding zone and with ejections from impacts of large bodies on the growing Earth.

Galimov and coauthors (Galimov et al., 2005; Galimov, 1995, 2008, 2011, 2013; Galimov and Krivtsov, 2012; Vasiliev et al., 2011) developed a model for the formation of the Earth and Moon embryos from a single initial rarefied gas and dust condensation in the protoplanetary disk, followed by the formation and compression of two fragments. This model satisfies geochemical constraints and also makes it possible to explain the known differences in the chemical composition of the Moon and the Earth, including iron deficiency, depletion of volatiles and enrichment in refractory oxides of Al, Ca, Ti in the substance of the Moon compared to the Earth. Galimov et al. suggested that in the inner part of the condensation, about 40% of the volatile matter (including FeO) of the dust particles that formed the nuclei evaporated, and the initially high-temperature nuclei of the Moon and Earth were equally depleted in iron. The nuclei grew by accumulating particles rich in iron, which were located in the outer part of the condensation at the time of formation of the embryos. The growth of the Earth embryo was faster than the growth of the Moon embryo.

According to the model presented in (Galimov, Krivtsov, 2012), a rarefied condensation with a mass equal to the mass of the Earth–Moon system, about 50–70 million years after the beginning of the formation of the Solar System, contracted in $10^4$–$10^5$ years, forming the embryos of the Earth and the Moon. Such a long period of existence of condensations after the beginning of the formation of the Solar System was not obtained in the papers of specialists studying the formation and evolution of condensations. In most works devoted to the formation of planetesimals (Cuzzi et al., 2008, 2010; Cuzzi and Hogan, 2012; Johansen et al., 2007, 2009a, 2009b, 2011, 2012; Lyra et al., 2008, 2009; Youdin, 2011; Youdin and Kenyon, 2013), the immediate time of formation (after the start of formation from a compacted gas–dust disk) and compression of rarefied condensations did not exceed 1000 revolutions around the Sun, and in some models it occurred in just a few tens of revolutions around the Sun. In our opinion, it is impossible to explain the fact that in Galimov's model the fall of particles from the outer part of the condensation to the center of the condensation (on the embryos of the Earth and the Moon) began 50 million years later, and not immediately after the formation of the condensation. Indeed, in their model, the outer part of the condensation was cold, and the dust grains had zero velocities. Criticism of Galimov's model was discussed in more detail in the paper (Ipatov, 2018).

Ipatov (2018) showed that the angular momentum of the condensation, necessary for the formation of the embryos of the Earth–Moon system, could be acquired in the collision of two condensations with a total mass greater than $0.01m_E$ ($m_E$ is the mass of the Earth). Most of the material that entered the forming Moon was ejected from the Earth during numerous collisions with planetesimals and smaller bodies. A similar mechanism was proposed by Ipatov (2017a, 2017b) for the model of the formation of trans-Neptunian satellite systems. Unlike the condensation that gave birth to the embryos of the Earth and the Moon, the condensation that gave birth to the embryo of Mars did not have a large angular momentum and, during compression, could only give birth to the small satellites Phobos and Deimos. The angular momenta of the condensations that gave birth to the embryos of Mercury and Venus were insufficient even for the formation of small satellites. Unlike Galimov's model, in Ipatov's model (2018), the Earth's embryo grew due to the accumulation of planetesimals on it, and not dust from the outer part of the condensation with a mass on the order of the Earth's mass. Unlike the multi-impact model, in Ipatov's model the initial embryo of the Moon was formed from the same condensation as the embryo of the Earth. Objects ejected from the Earth's embryo during collisions with other objects had a greater chance of becoming part of the large Moon's embryo than of combining with similar low-mass objects. The embryo of the Moon, formed during the compression of the condensation, could be the trigger for its growth to the present mass of the Moon, mainly due to the material ejected from the Earth during its bombardment by planetesimals. In our opinion, for the multi-impact model, all terrestrial planets would have large satellites, since the same bodies fell on them as on the Earth's embryo. To explain the present iron content in the Moon, Ipatov (2018) estimated that the fraction of matter ejected from the Earth's embryo and falling onto the Moon's embryo should have been almost an order of magnitude greater than the sum of the total mass of planetesimals that fell directly onto the Moon's embryo and the initial mass of the Moon's embryo, formed from the parent condensation, provided that this initial lunar embryo contained the same proportion of iron as the planetesimals.

According to (Salmon and Canup, 2012; Cuk and Stewart, 2012), the semimajor axis of the orbit of the formed Moon embryo was equal to $6r_E$–$7r_E$, where $r_E$ is the radius of the Earth. Due to tidal interactions, the distance between the Earth and the Moon could relatively quickly reach $30r_E$ (Touma and Wisdom, 1994; Pahlevan and Morbidelli, 2015). For example, Pahlevan and Morbidelli (2015) found that the distance between the Earth and the Moon increased to $20r_E$–$40r_E$ in $10^6$–$10^7$ years. The current distance between the Earth and the Moon is $60.4r_E$. In the multi-impact model it is considered (Rufu et al., 2017), that, due to tidal interactions, the resulting satellite of the Earth with a smaller mass moved away from the Earth over time and caught up with the previously formed satellite that was more massive and initially more distant from the Earth.

### Fallout of Bodies Ejected from the Earth onto other Planets and the Moon

In the papers (Gladman et al., 2005) and (Reyes-Ruiz et al., 2012) the evolution of the orbits of bodies ejected from the Earth was simulated over a period of 30 000 years, and in (Reyes-Ruiz et al., 2012) the number of bodies considered was 50 times greater than in (Gladman et al., 2005). These authors carried out calculations of the evolution of the orbits of matter ejected from the Earth, with multiple values of ejection velocity $v_{ej}$ from 11.22 to 16.4 km/s. In (Gladman et al., 2005), the probability of a collision of an ejected body with the Earth over 30 thousand years varied from 0.09 at $v_{ej}$ = 11.22 km/s up to 0.001–0.003 at $13.2 \le v_{ej} \le 16.4$ km/s. The probabilities of collisions of ejected bodies with the Earth, Venus and Mars were calculated. (Reyes-Ruiz et al., 2012) used the Mercury integration package (Chambers, 1999). As in (Gladman et al., 2005), the initial bodies were launched vertically, and the starting points were distributed evenly over the Earth's surface. The bodies started from a height $h$ = 100 km above the Earth. Integration ended when the ejected body collided with the Moon, planet or Sun, or moved away from the Sun by 40 AU, or if the time interval under consideration reached 30 thousand years. The probabilities of collisions of bodies with the Earth, Moon, Venus, Mars, Jupiter, Saturn and the Sun were calculated.

Armstrong et al. (2002) studied the fallout of material ejected from the Earth during the Late Heavy Bombardment. They believed that, according to Zharkov (2000), the semimajor axis of the Moon's orbit was $21.6r_E$ 3.9 billion years ago. It was believed that with the impactor's velocity equal to 14 km/s, the mass of matter ejected from the Earth was equal to 0.14 of the impactor's mass. The considered time interval after the ejection of the substance did not exceed 1000 years, while the majority of collisions (90%) occurred within a period of no more than 100 years. The fallout of bodies back to the Earth and the Moon occurred mainly at ejection velocities close to the parabolic velocity on the Earth's surface. From Table 2 in (Armstrong et al., 2002) it follows that at velocity $v_\infty$ of ejected bodies at infinity equal zero; the fractions of bodies falling on the Earth and the Moon were 0.61 and 0.004, respectively. Moreover, the ratio of these fractions was 152.5. At $v_\infty$ = 1 km/s these fractions were 0.06 and 0, respectively.

### Ejection of Matter from a Crater

Figure 12.3 of the monograph (Melosh, 1994) shows the ratio of the lost mass to the mass of the impactor depending on the impact velocity $v_{col}$ and the escape velocity $v_{esc}$ of the celestial object. The less $v_{col}$, the smaller this mass ratio. From this figure it is clear that for the Earth this ratio is close to 0.1 and 0.01 at $v_{col}$ equal to 45 and 30 km/s, respectively. For the Moon, this ratio is less than 0.1 and 0.01 at $v_{col}$, equal to 7.5 and 5 km/s respectively. It is close to 1 for $v_{col}$ close to 20 km/s.

**Table 1.** The ratio $k$ of probabilities of collisions with the Earth and the Moon at the semimajor axis $a_M$ (equal to 60 or 5 Earth radii $r_E$) of the Moon's orbit and the initial ratio $k_{pE/pM}$ of these probabilities (equal to 20, 30 or 40) at a distance equal to the radius $a_{Ms}r_E$ of the Hill sphere ($a_{Ms} = 230.7$) or the sphere of action of the Earth ($a_{Ms} = 145.8$) for the model under consideration

|  | $a_{Ms} = 230.7$ | $a_{Ms} = 145.8$ | $a_{Ms} = 230.7$ | $a_{Ms} = 145.8$ |
|---|---|---|---|---|
|  | $a_M = 60$ | $a_M = 60$ | $a_M = 5$ | $a_M = 5$ |
| $k_{pE/pM} = 20$ | 16.5 | 17.4 | 13.96 | 14.1 |
| $k_{pE/pM} = 30$ | 21.2 | 23.45 | 14.8 | 15.2 |
| $k_{pE/pM} = 40$ | 25.9 | 29.5 | 15.7 | 16.35 |

Artemieva and Shuvalov (2008) numerically simulated impacts on the Moon by asteroids with a velocity of 18 km/s and icy bodies with velocities of 25 and 55 km/s at angles from 15° to 90° to the surface. Maximum ejections with velocities higher than the escape velocity of the Moon were achieved at angles of 45°–60° and ranged from 4 to 12 masses of the impacting body upon impact of an asteroid and a parabolic comet, respectively. These estimates were several times larger than the approximate estimates in Fig. 12.3 in (Melosh, 1994). The mass loss of the Moon due to impacts of cosmic bodies was estimated.

Artemieva (2014) using numerical simulation compared impacts of asteroids 1–500 m in size with a velocity of 18 km/s at an angle of 45° on a porous medium simulating lunar regolith, with impacts on a solid nonporous target. It turned out that porosity can reduce the ejection of solid fragments, which are the source of lunar meteorites, by an order of magnitude. The number of bodies falling onto the front hemisphere of the lunar surface is slightly greater than the number of bodies falling onto the rear hemisphere (Gallant et al., 2009; Wang and Zhou, 2016).

According to formula (4) from (Armstrong et al., 2002), the mass of bodies ejected at velocities greater than $v_{ej}$, is equal to $m_{ej} = 0.11(\rho/\rho_t)^{0.2}(v/v_{ej})^{1.2}m_i$, where $\rho$ and $\rho_t$ are the density of the impactor and target, $v_i$ is the collision velocity, and $m_i$ is the impactor mass. At the same densities, $v_i = 14$ km/s and $v_{ej} = v_{esc}$ Armstrong et al. (2002) obtained the mass of bodies ejected from the Earth equal to $0.14m_i$. They noted that ejection velocities are generally less than $0.85v_i$.

Svetsov (2011) obtained a number of other dependences $m_{ej}$ on $v_i/v_{ej}$: (1) $m_{ej} = 0.03(\rho/\rho_t)^{0.2}(v_i/v_{ej})^{1.65}m_i$ based on experiments with basalts at velocities of 6 km/s given in (Holsapple and Housen, 2007) and (2) $m_{ej} = 0.03(v_i/v_{ej})^{2.3}m_i$ based on their numerical calculations at $v_i < 10$ km/s. According to these formulas, when $v_i/v_{ej}$ equal to 1 we get $m_{ej} = 0.03m_i$, and when $v_i/v_{ej} = 2$ we have $m_{ej} = 0.094m_i$ and $m_{ej} = 0.148m_i$, respectively, for the first and second formulas above. According to (Svetsov, 2011), the mass $m_{ej}$ of ejected material is greater than the mass of the impactor, if the collision velocity is not less than 3–5 times the parabolic velocity on the target surface. Svetsov et al. (2012) noted that the mass of matter ejected from the Earth was maximum at collision angles of about 30°–45° and did not exceed 0.04, 0.06, and 0.13 from the mass of the impactor at collision velocities equal to 12, 15, and 20 km/s, respectively.

According to formula (7.12.3) from (Melosh, 1994) $m_{ej}$ is proportional to $v_{ej}^{-1.7}$ and $v_{ej}^{-1.2}$ for water and sand, respectively. According to Fig. 12.3 from this work, the mass of material ejected from the Earth was approximately 0.01 and 0.1 of the impactor mass at impact velocities of 30 and 45 km/s, respectively. For the Moon, the mass of the ejected material exceeded the mass of the impactor at approximate velocities greater than 20 km/s. Based on Figs. 3 and 5 from (Shuvalov and Trubetskaya, 2011) we can conclude that at a collision velocity of 18 km/s, the ratio of the mass of the ejected substance (at velocities greater than the parabolic velocity on the Earth's surface) to the mass of the impactor was of the order of 0.002 and 0.2 for vertical and oblique (at 45°) impacts, respectively. In these calculations, the impactor and target were granite.

The results of the above works indicate that when the body collides with the Earth at a velocity of no more than 20 km/s, the mass of the ejected substance does not exceed 0.15 of the mass of the impactor.

According to (Shuvalov and Trubetskaya, 2011), values of an ejection angle $i_{ej}$ are mainly between 20° and 55°, especially between 40° and 50°. Close angle values were obtained in (Raducan et al., 2022) when bodies fall onto asteroids. In experiments with micron particles with velocities up to 2.5 km/s in (Barnouin et al., 2019), ejection angles were obtained from 40° to 80° at velocities of about 1 km/s and from 40° to 70° at higher velocities. At the maximum velocity considered (2.5 km/s), this range was even narrower: from 43° to 59°.

Beech et al. (2019) noted that due to the atmosphere, only bodies with a diameter of at least 0.7 m can leave the Earth. They could be ejected during the formation of a crater with a diameter of at least 10 km. The youngest terrestrial craters of this size formed about a million years ago.

Marov and Ipatov (2021) discussed the velocity of collisions of bodies with the Earth and the Moon. The collision velocities of planetesimals formed in the Earth's feeding zone with the Earth and Moon of present masses were approximately 13–15 and 8–10 km/s, respectively. For planetesimals from other feeding zones of terrestrial planets, these velocity ranges were equal to 13–19 and 8–16 km/s. As noted above, at collision velocities less than 20 km/s, the fraction of matter ejected from the Earth probably did not exceed 0.15. For the majority of bodies coming to the Earth from the zone of the outer asteroid belt and the zone of giant planets, these ranges were 23–26 and 20–23 km/s for the Earth and Moon, respectively. However, for individual bodies these velocities could be close to 40 km/s. As noted above, the velocities of most bodies ejected from the Earth were less than the collision velocities.

## MODELS CONSIDERED, CALCULATION VARIANTS AND INITIAL DATA

This paper examines the evolution of the orbits of bodies ejected from the Earth. In each variant of the calculations, the motion of 250 bodies ejected from the Earth was studied at fixed values of the ejection angle $i_{ej}$ (measured from the surface plane or from another parallel plane), ejection velocity $v_{ej}$ and time integration step $t_s$. In different variants, the values of the ejection angle $i_{ej}$ were 15°, 30°, 45°, 60°, 89° or 90°. The velocity $v_{ej}$ of bodies ejected from the Earth was generally equal to 11.22, 11.5, 12, 14, 16.4 or 20 km/s. Ejection of bodies at $v_{ej}$ = 20 km/s are quite rare, but are possible during high-velocity collisions with the Earth of bodies coming from beyond the orbit of Jupiter. Other velocity values were also considered in the range from 11.22 to 11.5 km/s, since in this range the dependence of the probabilities of collisions of bodies with the Moon and planets on $v_{ej}$ is stronger than for higher values of $v_{ej}$. The velocity of rotation of the Earth's surface around the Earth's axis (0.46 km/s) and the velocity of the Moon around the Earth (1.02 km/s) were not taken into account. They are small compared to the ejection velocities considered.

A symplectic algorithm was used to integrate the equations of motion from the SWIFT package (Levison and Duncan, 1994). The gravitational influence of the Sun and all eight planets was taken into account. The initial positions of the planets in their orbits were taken from the test file of this package. Bodies that collided with planets or the Sun or reached 2000 AU from the Sun were excluded from integration. In the considered variants of calculations of the motion of bodies ejected from the Earth, most of the bodies left the Hill sphere of the Earth and moved in heliocentric orbits. The motion of bodies ejected from the Earth was studied during the maximum dynamic lifetime $T_{end}$ of all bodies considered in the variant, which in the calculation variants was often about 200–700 million years. Such ejection often occurred during the Earth's accumulation and late heavy bombardment stages. Previously (Gladman et al., 2005; Reyes-Ruiz et al., 2012), when modeling the ejection of bodies from the entire Earth's surface, initial velocities perpendicular to the Earth's surface and a time interval of 30 thousand years were considered. In reality, as noted in the Introduction, ejection angles generally do not exceed 55°, and most of the bodies ejected into hyperbolic orbits remain in orbits after 30 thousand years. The probability of a body falling onto the Earth is different for different points on the Earth's surface (for example, from the space beyond the Earth's orbit it is greater than from the space inside the Earth's orbit).

The considered time integration step $t_s$ in some variants was equal to 1, 2, 5 or 10 days, and the calculation results obtained for different $t_s$ have been compared. These calculations gave similar results. Unless otherwise stated, the paper provides data for calculations with $t_s$ equal to five days. In (Frantseva et al., 2022) it was noted that in the considered symplectic algorithm from the SWIFT package, the time integration step decreases significantly when the distance between objects is less than 3.5 Hill radii. The calculation results given in (Ipatov and Mather, 2004a; 2004b), showed that the symplectic algorithm and the BULSTO algorithm (Bulirsh and Stoer, 1966) give similar results when studying the motion of bodies in the Solar System.

The ejection of bodies from six opposite points on the Earth's surface was considered for a number of values of velocities and angles of ejection of bodies. In the series of calculations *vf* and *vc*, the motion of the bodies began, respectively, from the points most and least distant from the Sun *F* (far) and *C* (close) of the Earth's surface on the line from the Sun to the Earth. In the series *vw* and *vb*, the bodies started from the points *W* (forward) and *B* (back), respectively, of the surface of the Earth in the direction of the Earth's motion and from the opposite side of the Earth. In series *vu* and *vd*, the bodies started, respectively, from the points *U* (up) and *D* (down) of the Earth's surface with maximum and minimum values *z* (with axis *oz*, perpendicular to the plane of the Earth's orbit). In the series *vf* for several initial data, the bodies started from a height *h* above the surface of the Earth. Unless otherwise stated, below, we are talking about bodies launched from the surface of the Earth (at $h = 0$).

The probabilities of collisions of ejected bodies with the Earth and other planets were calculated by integrating the equations of motion as the number of bodies colliding with the planet to the initial number of bodies. Moreover, the probabilities of collisions of bodies with the Earth and the Moon were calculated based on arrays of orbital elements of migrated bodies (stored in increments of 500 years for the entire time of orbital evolution) similarly to (Ipatov and Mather, 2003; 2004a; 2004b; Ipatov, 2000; 2019). The basic algorithm for calculating probability has been described by Ipatov (1988; 2000), and an addition to the algorithm that takes into account the change in the velocity of an object moving in an eccentric orbit is given in (Ipatov and Mather, 2004b). In this algorithm, bodies were considered to move around the Sun in unperturbed Keplerian orbits outside the sphere of action of the larger body. Motion within the sphere of action was considered within the framework of the two-body problem. The algorithm used is more complex than the commonly used Őpik approach (Őpik, 1951), and the calculated probability of approach to the sphere of action also depends on the synodic period of revolution of the two bodies around the Sun.

The values of $k_{pE/pM} = p_E^*/p_M^*$ for the probability ratios $p_E^*$ and $p_M^*$ of collisions of bodies with the Earth and the Moon, obtained on the basis of arrays of orbital elements, were also calculated. This ratio was calculated only for bodies that initially left the Earth's sphere of action. At low ejection velocities, most of the bodies ejected from the Earth could fall back to the Earth without leaving the Earth's sphere of action. Below, the probability of a body colliding with the Moon was calculated using the formula $p_M = p_E/k_{pE/pM}$, where $p_E$ is the probability of a body colliding with the Earth, calculated by integrating the equations of motion (regardless of whether the body left the sphere of action of the Earth).

Values of $p_E^*$ and $p_M^*$ were calculated under the assumption that the Earth and the Moon move independently around the Sun in the same heliocentric orbit, i.e., the Moon is outside the Hill sphere of the Earth. In reality, the ratio of the probabilities of collisions of bodies with the Earth and the Moon may be less than $k_{pE/pM}$ and smaller the closer the Moon's orbit is to the Earth. As noted in the next section, in the calculations of the value of $k_{pE/pM}$, it could often be around 30. If, hypothetically, the Moon was located at the very surface of the Earth, then this ratio of collision probabilities would be equal to the ratio of the squares of the radii of the Earth and the Moon, i.e., 13.39. Therefore, depending on the semimajor axis of the Moon's orbit around the Earth, this ratio can range from 13.4 to $k_{pE/pM}$. For bodies that entered the Hill sphere and could then fall to the Earth, there was a chance of colliding with the Moon if it fell into the cone of trajectories that reached the Earth's surface. This cone is limited by parabolic trajectories that reach the Earth. Therefore, it can be approximately assumed that the probability $k$ of collisions with the Earth and the Moon depend on the semimajor axis $a_M$ (in Earth radii $r_E$) of the Moon's orbit as $k = ba_M^{1/2} + c$. Believing that $k = k_{pE/pM}$ with $a_M = a_{Ms} = 230.7$ (at the boundary of the Earth's Hill sphere) and $k = 13.4$ with $a_M = 1$, we obtain $b = 0.07k_{pE/pM} - 0.944$ and $c = 13.4 - b = 14.344 - 0.07k_{pE/pM}$. When calculating $k_{pE/pM}$ the method of spheres of action was used. The radius of the Earth's sphere of action is 929 000 km, i.e., $a_M = a_{Ms} = 145.8$ times the radius $r_E$ of the Earth. With this $a_M = 145.8$ we have $b = 0.09k_{pE/pM} - 1.21$ and $c = 14.61 - 0.09k_{pE/pM}$.

Table 1 shows the values of the ratio $k$ of the probabilities of collisions with the Earth and the Moon for the ratio $a_M$ of the semimajor axis of the Moon's orbit to the radius of the Earth $r_E$, equal to 60 or 5, and at this initial ratio $k_{pE/pM}$ at a distance equal to the radius $a_{Ms}r_E$ of the Hill sphere ($a_{Ms} = 230.7$) or the sphere of action of the Earth ($a_{Ms} = 145.8$). For the two considered values of $a_{Ms}$ (230.7 and 145.8) the values of $k$ are quite close. The values of $k$ are less for smaller values of the semimajor axis of the Moon's orbit. With $a_M = 5$ the values of $k$ depended little on $k_{pE/pM}$ and were about 14–16. With $a_M = 60$, the ratio $k/k_{pE/pM}$ was in the ranges 0.825–0.87, 0.71–0.78 and 0.65–0.74 with $k_{pE/pM}$ equal to 20, 30, and 40, respectively. At large inclinations of the orbits of bodies within the sphere of action of the Earth, the Moon's orbit may not intersect the above mentioned cone of trajectories of bodies moving towards the Earth, or it may intersect the cone not at its center. Therefore, the real ratio of the probability of collisions of bodies with the Earth and the Moon may be greater than the values of $k$ given in Table 1. However, it does not exceed $k_{pE/pM}$.

Below, in the next section it is noted that the velocities of entry of bodies into the sphere of action of the Earth after their initial ejection from it are mainly in the range of 10–25 km/s. This velocity increases as the body approaches the Earth. The parabolic velocity on the surface of the Moon is 2.38 km/s. For a relative velocity of 10 km/s, the effective radius of the Moon is only 1.028 times greater than its physical radius, that is, both radii are almost the same. At higher relative velocities, the effective radius of the Moon is smaller. The square of the effective radius $r_{eff}$ of a celestial body radius $r$ equals $r_{eff}^2 = r^2[1 + (v_{par}/v_{rel})^2]$, where $v_{par}$ is the parabolic velocity on the surface of this celestial body, and $v_{rel}$ is the relative velocity of a small body when it enters the sphere of action of a celestial object (the exact formula is valid for the relative velocity at infinity).

On the basis of the ratio $k_{pE/pM}$ of the number of planetesimals colliding with the Earth and the Moon, we can estimate (Marov and Ipatov, 2021) the characteristic velocities $v_{relE}$ (relative to the Earth) of planetesimals upon entering the Earth's sphere of action according to the formula: $(v_{relE}/v_{parE})^2 = [k_{pE/pM}(v_{parM}/v_{parE})^2(r_M/r_E)^2 - 1]/[1 - k_{pE/pM}(r_M/r_E)^2]$, where

$r_M$ and $r_E$ are the radii of the Moon and Earth, and $v_{parM}$ = 2.38 km/s and $v_{parE}$ = 11.186 km/s are the parabolic velocities near the surface of the Moon and Earth, respectively. The ratio $k_{pE/pM}$ was calculated on the basis of the elements of the orbits of bodies over a large time interval, during which many approaches of bodies with the Earth could occur up to the radius of its sphere of action. Therefore, the above formula allows us to obtain the characteristic velocities of bodies entered in the sphere of action. The velocities of different approaches could be different and could differ from these characteristic velocities. For velocities $v_{colE}$ and $v_{colM}$ of collisions of bodies with the Earth and the Moon, the following relations are valid: $v_{colE}^2 = v_{relE}^2 + v_{parE}^2$ and $v_{colM}^2 = v_{relM}^2 + v_{parM}^2$. When $v_{relE} = v_{relM}$, based on the $v_{relE}$ values, we also calculated the values of $v_{colM}$ and $v_{colE}$.

The results of the calculations (with $h = 0$ and $t_s = 5^d$) are presented in Figs. 1–3. Figure 1 shows the values $k_{pE/pM} = p_E^* / p_M^*$ of the ratios of the probabilities of collisions of bodies with the Earth and the Moon, obtained on the basis of arrays of orbital elements. Figure 2 shows the probability of collisions of bodies with the Moon, calculated using the formula $p_M = p_E/k_{pE/pM}$. Figure 3 shows the values of the collision velocities of bodies with the Moon and the Earth. The values presented in Figs. 1–3 are given depending on the ejection angle $i_{ej}$ for different values of ejection velocity $v_{ej}$ and for six ejection points from the Earth's surface. For a small number of variants (with fixed values of $v_{esc}$, $i_{ej}$ and ejection points), two variants of calculations were carried out, each with 250 initial bodies. In this case, the figures show two identical icons with the same value of $i_{ej}$.

## PROBABILITIES AND VELOCITIES OF COLLISIONS OF EJECTED BODIES WITH THE MOON

With ejection velocities $v_{ej}$ no less than 11.3 km/s, most of the ejected bodies left the Hill sphere of the Earth and began to move around the Sun. During evolution, they could collide with planets or the Sun or be thrown into hyperbolic orbits (Ipatov, 2023). In this work, we focus on studying the fallout of ejected bodies on the Moon.

In the previous section it was noted that the probability of collisions of bodies with the Moon was calculated using the formula $p_M = p_E/k_{pE/pM}$, where $p_E$ is the probability of a collision with the Earth obtained by integrating the equations of motion, and $k_{pE/pM} = p_E^* / p_M^*$ is the ratio of the probabilities of collisions of bodies with the Earth and the Moon, obtained on the basis of arrays of orbital elements. Therefore, we first focus on the values $p_E$ and $k_{pE/pM}$.

The results show that at higher ejection velocities $v_{ej}$ the probabilities $p_E$ and $p_M$ of collisions of bodies with the Earth and the Moon were generally lower. The probabilities $p_E$ of collisions of bodies with the Earth were approximately the same for calculations with different considered values of the integration step $t_s$. At velocities $11.5 \leq v_{ej} \leq 14$ km/s the probability $p_E$ did not greatly depend on the ejection point on the Earth's surface. Below are the values $p_E$, obtained in variants with fixed values of $v_{ej}$ and $i_{ej}$. The average value of $p_E$ over all ejected bodies depends on the distribution of bodies over $v_{ej}$ and $i_{ej}$.

At ejection velocities $v_{ej} \leq 11.25$ km/s, i.e., slightly exceeding the parabolic velocity, most of the ejected bodies quickly fell back to the Earth. At $v_{ej}$ = 11.22 and $v_{ej}$ = 11.25 km/s some of the ejected bodies left the Earth's Hill sphere for only two ejection points (W and B) in opposite directions of the Earth's line of motion along the heliocentric orbit. For these two ejection points, the data in Figs. 1–3 are not given for the values $i_{ej}$, in which all bodies fell onto the Earth without leaving its sphere of action. The ejection of bodies from the sphere of action of the Earth at such velocities ($v_{ej}$ = 11.22 and $v_{ej}$ = 11.25 km/s) for other ejection points is possible only if the initial launch point is above the Earth's surface. For example, in the *vf* series with $v_{ej}$ = 11.22 km/s and $i_{ej} \geq 45°$, the evolution times of the orbits of the ejected bodies were 15 days, 25 days, 20 years and about 400 million years at the launch altitude $h$ above the Earth's surface, equal to 0, 20, 50 and 100 km, respectively. In these calculations at $h$ = 100 km and $i_{ej}$ = 45°, about 38%, 44 and 49% of the initial bodies collided with the Earth in times equal to 1, 10 and $T_{end}$ = 393 million years, respectively.

In Figs. 1–3, data are given for the finite time intervals considered $T = T_{end}$ (when all bodies have already collided with planets or the Sun or have reached 2000 AU from the Sun), which, when $v_{ej} \geq 11.3$ km/s were mainly located in the interval from 200 to 700 Ma and reached 1356 Ma. For such $v_{ej}$ values $T_{end}$ < 100 Ma were obtained only in the series *vw* of calculations at $v_{ej}$ = 16.4 km/s and $i_{ej} \geq 60°$, and also when $v_{ej}$ = 20 km/s and $i_{ej} \geq 45°$.

In Table 2 with $i_{ej}$ = 45° the values of the ratio of the probabilities of collisions of bodies with the Earth for five time intervals from 0.01 to 100 million years to similar probabilities at the end of evolution are given for several series $N_v$ of calculations (for different ejection points) and several ejection velocities $v_{ej}$. To estimate the probabilities of collisions of bodies with the Earth and the Moon during these times, these ratios need to be multiplied by the probabilities $p_E$ and $p_M$ of collisions of bodies with the Earth and the Moon, respectively, over the entire considered time interval. These probabilities are also given in Table 2. Table 3 shows the intervals of the values $p_E$ of probabilities of

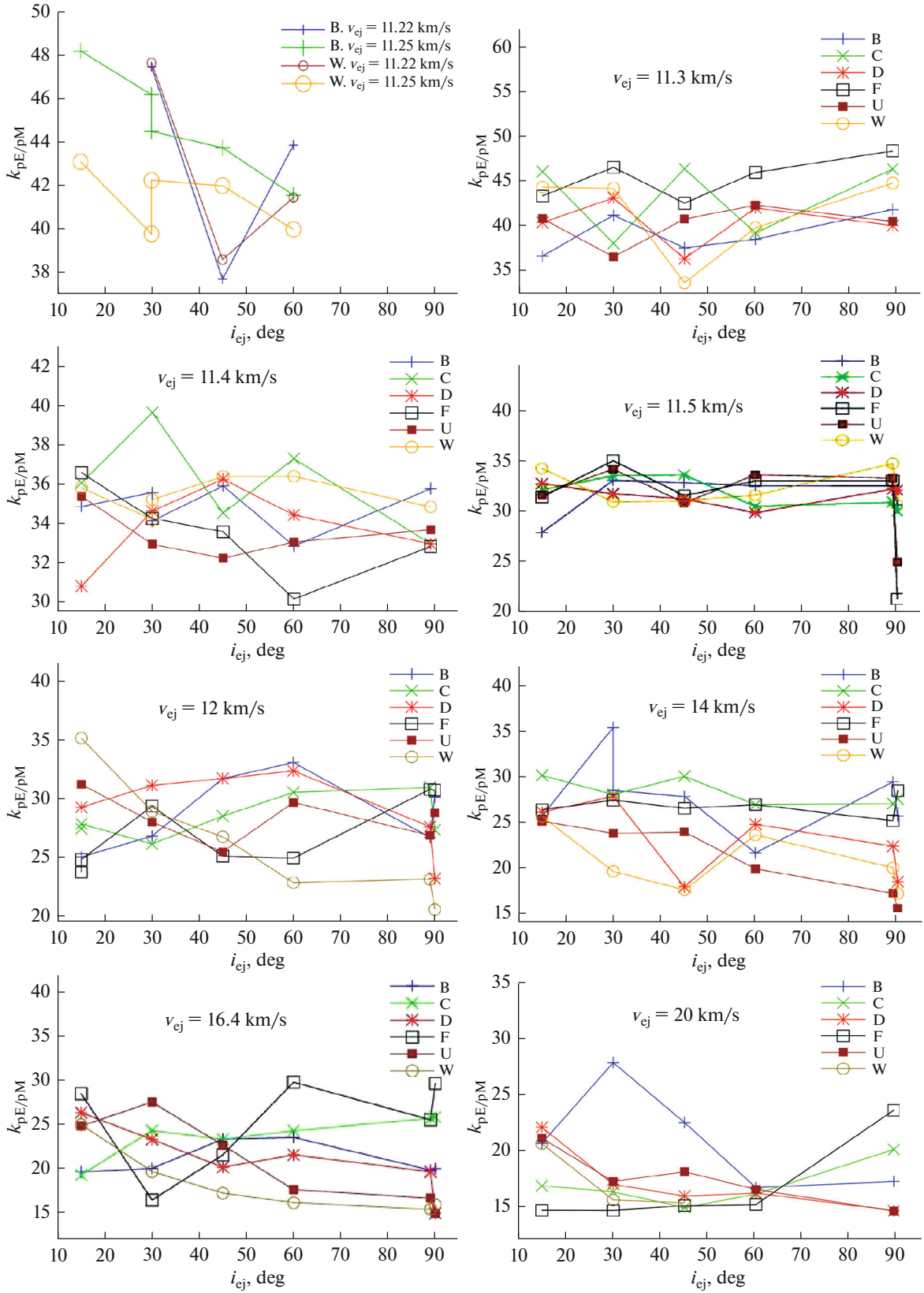

**Fig. 1.** The values $k_{pE/pM} = p_E^*/p_M^*$ of the ratio of the probabilities of collisions of bodies with the Earth and the Moon, obtained on the basis of arrays of orbital elements, depending on the ejection angle $i_{ej}$ (in degrees) for a range of values of ejection velocity $v_{ej}$ (in km/s) and various ejection points.

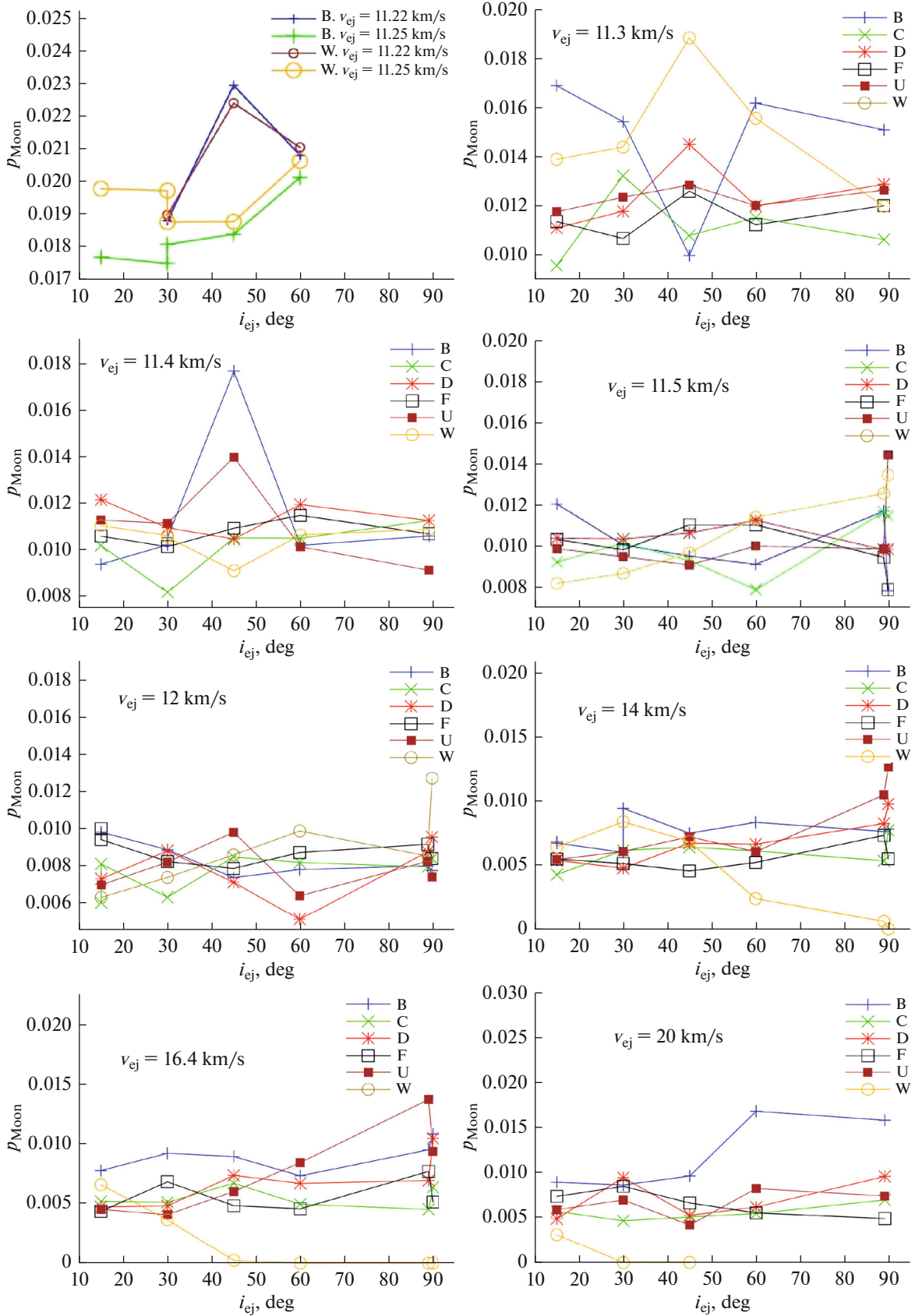

**Fig. 2.** The probability of collisions of bodies with the Moon, calculated using the formula $p_M = p_E/k_{pE/pM}$, depending on the ejection angle $i_{ej}$ (in degrees) for a range of values of the ejection velocity $v_{ej}$ (in km/s) and various ejection points.

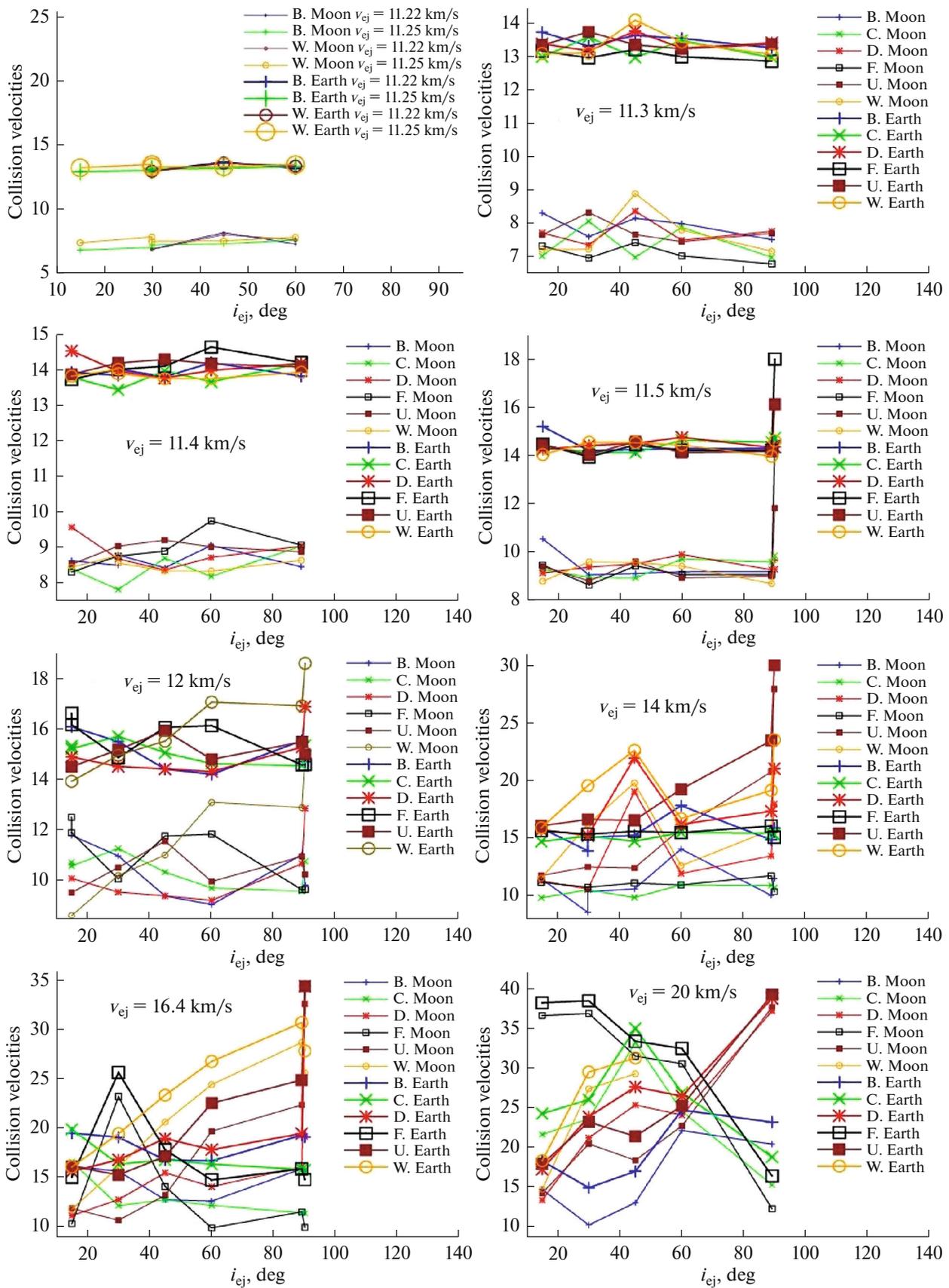

**Fig. 3.** Values of collision velocities (in km/s) of bodies with the Moon and Earth depending on the ejection angle $i_{ej}$ (in degrees) for a range of values of ejection velocity $v_{ej}$ (in km/s) and various ejection points.

**Table 2.** Probabilities of collisions of bodies with the Earth and the Moon at different time

| $p$ | $N_v$ | $v_{ej}$, km/s | | | | | $N_v$ | $v_{ej}$, km/s | | | | |
|---|---|---|---|---|---|---|---|---|---|---|---|---|
| | | 11.3 | 11.5 | 12 | 14 | 16.4 | | 11.3 | 11.5 | 12 | 14 | 16.4 |
| $p_{001}$ | vw | 0.642 | 0.053 | 0.018 | 0 | 0 | vb | 0.647 | 0.026 | 0.017 | 0.0 | 0.0 |
| $p_{01}$ | vw | 0.774 | 0.224 | 0.035 | 0.033 | 0 | vb | 0.760 | 0.269 | 0.052 | 0.019 | 0.019 |
| $p_1$ | vw | 0.899 | 0.547 | 0.298 | 0.167 | 0 | vb | 0.856 | 0.526 | 0.276 | 0.173 | 0.115 |
| $p_{10}$ | vw | 0.937 | 0.827 | 0.667 | 0.333 | 0 | vb | 0.916 | 0.782 | 0.707 | 0.558 | 0.519 |
| $p_{100}$ | vw | 1.0 | 0.987 | 1.0 | 0.90 | 1.0 | vb | 1.0 | 0.987 | 0.931 | 0.962 | 0.962 |
| $p_E$ | vw | 0.636 | 0.30 | 0.228 | 0.156 | 0.004 | vb | 0.668 | 0.312 | 0.232 | 0.208 | 0.208 |
| $p_M$ | vw | 0.019 | 0.010 | 0.009 | 0.007 | 0.0002 | vb | 0.018 | 0.010 | 0.007 | 0.007 | 0.009 |
| $p_{001}$ | vf | 0.209 | 0.046 | 0.041 | 0.033 | 0.038 | vc | 0.144 | 0.064 | 0.017 | 0.021 | 0.0 |
| $p_{01}$ | vf | 0.425 | 0.218 | 0.041 | 0.033 | 0.038 | vc | 0.496 | 0.141 | 0.083 | 0.042 | 0.026 |
| $p_1$ | vf | 0.724 | 0.552 | 0.306 | 0.133 | 0.192 | vc | 0.728 | 0.462 | 0.30 | 0.146 | 0.077 |
| $p_{10}$ | vf | 0.918 | 0.816 | 0.816 | 0.60 | 0.577 | vc | 0.912 | 0.679 | 0.633 | 0.562 | 0.590 |
| $p_{100}$ | vf | 0.985 | 0.977 | 0.918 | 0.966 | 1.0 | vc | 0.992 | 0.949 | 0.950 | 0.938 | 0.923 |
| $p_E$ | vf | 0.536 | 0.348 | 0.196 | 0.120 | 0.104 | vc | 0.50 | 0.312 | 0.240 | 0.192 | 0.156 |
| $p_M$ | vf | 0.013 | 0.011 | 0.008 | 0.005 | 0.005 | vc | 0.011 | 0.009 | 0.008 | 0.006 | 0.007 |
| $p_{001}$ | vu | 0.328 | 0.029 | 0.016 | 0.0 | 0.0 | vd | 0.303 | 0.060 | 0.032 | 0.0 | 0.0 |
| $p_{01}$ | vu | 0.580 | 0.186 | 0.048 | 0.047 | 0.0 | vd | 0.636 | 0.277 | 0.032 | 0.033 | 0.0 |
| $p_1$ | vu | 0.815 | 0.471 | 0.274 | 0.116 | 0.088 | vd | 0.856 | 0.518 | 0.238 | 0.066 | 0.054 |
| $p_{10}$ | vu | 0.885 | 0.714 | 0.613 | 0.558 | 0.559 | vd | 0.947 | 0.783 | 0.635 | 0.533 | 0.297 |
| $p_{100}$ | vu | 0.992 | 0.929 | 0.968 | 0.907 | 0.971 | vd | 1.0 | 0.952 | 0.968 | 0.933 | 0.919 |
| $p_E$ | vu | 0.524 | 0.280 | 0.248 | 0.172 | 0.136 | vd | 0.528 | 0.332 | 0.224 | 0.120 | 0.148 |
| $p_M$ | vu | 0.013 | 0.009 | 0.010 | 0.007 | 0.006 | vd | 0.007 | 0.011 | 0.007 | 0.007 | 0.007 |

Notes: Ratios $p_{001}$, $p_{01}$, $p_1$, $p_{10}$ and $p_{100}$ of probabilities of collisions of bodies with the Earth over time $T$, equal to 0.01, 0.1, 1, 10 and 100 million years, respectively, to similar probabilities at the end of evolution for several series $N_v$ of calculations (vw, vb, vf, vc, vu and vd) and several velocities $v_{ej}$ (in km/s) of ejection (given in the first row of the table) at ejection angle $i_{ej}$ equal to 45°. $p_E$ and $p_M$ are the probabilities of collisions of bodies with the Earth and the Moon, respectively, over the entire considered time interval. Values of $p_{001}$, $p_{01}$, $p_1$, $p_{10}$, $p_{100}$, $p_E$ and $p_M$ are given in the rows of the table, starting from the second row, and each row contains data for two ejection points (for two values of $N_v$). Data for different ejection points are separated by double lines

**Table 3.** Interval of the values $p_E$ of probabilities of collisions of bodies with the Earth over several time intervals $T$ (in Ma) for several series $N_v$ of calculations and several ejection velocities $v_{ej}$ (in km/s) at an ejection angle $i_{ej}$ equal to 45°

| $T$ | $v_{ej} = 11.3$ | $v_{ej} = 11.5$ | $v_{ej} = 12$ | $v_{ej} = 14$ | $v_{ej} = 16.4$ | $v_{ej} = 20$ |
|---|---|---|---|---|---|---|
| 0.01 | 0.072–0.432 | 0.008–0.02 | 0.004–0.008 | 0.0–0.004 | 0.0–0.004 | 0.0 |
| 0.1 | 0.228–0.508 | 0.044–0.092 | 0.008–0.02 | 0.004–0.008 | 0.0–0.004 | 0.0 |
| 1 | 0.364–0.572 | 0.132–0.192 | 0.06–0.072 | 0.008–0.036 | 0.0–0.024 | 0.0–0.02 |
| 10 | 0.456–0.612 | 0.2–0.284 | 0.152–0.164 | 0.04–0.116 | 0.0–0.092 | 0.0–0.096 |
| 100 | 0.496–0.668 | 0.26–0.34 | 0.18–0.244 | 0.108–0.20 | 0.004–0.2 | 0.0–0.192 |
| $T_{end}$ | 0.50–0.668 | 0.28–0.348 | 0.196–0.248 | 0.12–0.208 | 0.004–0.208 | 0.0–0.216 |

collisions of bodies with the Earth over several time intervals $T$ for several series $N_v$ of calculations and several emission velocities $v_{ej}$ (in km/s) at an ejection angle $i_{ej}$ equal to 45°.

At $v_{ej} \geq 12$ km/s, the probability of collisions of bodies with the Earth during the first 30 thousand years did not exceed 0.004, as in (Gladman, 2005; Reyes-Ruiz et al., 2012). The probabilities of collisions of bodies with the Earth are generally lower for larger values of $v_{ej}$. The dependence of the probability of collisions of bodies with the Moon on $v_{ej}$ is slightly less than with the Earth, since at high velocities when a body enters the Earth's sphere of action, the difference between the probabilities of collisions of bodies with the Earth and the Moon is less. The difference between probabilities $p_E$ of collisions of bodies with the Earth at different ejection points are greater for large values of $v_{ej}$. During $T = 1$ million years, $p_E$ in Table 3 could take values greater than 0.5 when $v_{ej} = 11.3$ km/s and be equal to zero at $v_{ej} \geq 16.4$ km/s.

In the considered variants for $i_{ej} = 45°$, the value of the ratio $p_E$ at $T = 10$ million years and at $T = 1$ million years did not exceed two for $v_{ej} \leq 11.5$ km/s and could reach four at higher values of $v_{ej}$. With $T \geq 10$ Ma and $v_{ej} \leq 12$ km/s, the values of $p_E$ for different ejection points and ejection angles $i_{ej} \leq 89°$ usually differed by no more than a factor of two. At $T = T_{end}$ and $v_{ej}$ equal to 11.5, 12, and 14 km/s, values of $p_E$ were approximately 0.3, 0.2, and 0.15–0.2, respectively. The only essential difference was in the variant $vw$ (for ejection from the forward point of the Earth's motion) at $v_{ej} \geq 16.4$ km/s. For the ejection point $W$ at $v_{ej} = 16.4$ km/s and $30° \leq i_{ej} \leq 60°$ at least 80% of the bodies were thrown into hyperbolic orbits and at least 17% of the bodies collided with the Sun. All bodies thrown from the point $W$ with $v_{ej} = 20$ km/s, were ejected into hyperbolic orbits at $i_{ej} \geq 45°$ (and, with $i_{ej} \geq 60°$, for a time less than 700 years), and the values of $p_E$ were equal to 0 and 0.064 at $i_{ej}$, equal to 30° and 15°, respectively. In this case, more than 90% of bodies were thrown into hyperbolic orbits at $i_{ej} = 30°$, and 70% of bodies fell into the Sun at $i_{ej} = 15°$. In some variants the values $p_E$ at $i_{ej} = 90°$ could even differ from the values $p_E$ at $i_{ej} = 89°$. For example, for the series $vf$ and $v_{ej} = 11.5$ km/s, the values of $p_E$ were equal to 0.31 and 0.17 at $i_{ej} = 89°$ and $i_{ej} = 90°$ respectively. These differences may be due to the fact that the original data at $i_{ej} = 90°$ were very close for different bodies in the same variant, and there was no such averaging along the ejection directions as with lower values $i_{ej}$. Values of ejection velocities at $i_{ej} = 90°$ could differ only in the ninth significant digit, but the coordinates were completely the same.

In Table 2 for $v_{ej} = 11.3$ km/s, the ratio of the number of bodies colliding with the Earth at some time $T$ to a similar number with $T = T_{end}$ is less different for pairs of the variants $vw$ and $vb$, $vf$ and $vc$, $vu$ and $vd$. Although in almost all variants of Table 2 at $v_{ej} \geq 11.3$ km/s the dynamical lifetimes of the last long-lived bodies exceeded 100 million years, more than half of the bodies that fell on the Earth fell on it during $T < 1$ Ma at $v_{ej} = 11.3$ km/s and for time $T < 10$ Ma at $v_{ej} \leq 12$ km/s. At $T = 100$ million years values of $p_E$ exceeded 90% of the values $p_E$ at $T = T_{end}$. Average value of $p_E$ depends on the distribution of bodies over $v_{ej}$ and $i_{ej}$. At $T = T_{end}$ it's probably around 0.2. The ratio of the probability of collisions of bodies with the Earth to the probabilities of collisions of bodies with other planets and the Sun usually decreased over time. The total number of bodies delivered to Earth and Venus was probably not much different.

**Values of the ratio $k_{pE/pM} = p_E^* / p_M^*$ of probabilities of collisions of bodies with the Earth and the Moon**, obtained based on arrays of orbital elements, are shown in Fig. 1 for the final time intervals considered. From this figure it can be seen that the values $k_{pE/pM}$ are mainly in the ranges 35–48, 30–40, 30–35, 25–32, 17–30, 15–30 and 15–25 at $v_{ej}$ equal to 11.3, 11.4, 11.5, 12, 14, 16.4 and 20 km/s respectively. A narrower range was obtained with $v_{ej} = 11.5$ km/s. On average, for different variants there was no significant difference in the values $k_{pE/pM}$ at different values $i_{ej}$ from 15° to 89°. Significant difference in values $k_{pE/pM}$ sometimes was with $i_{ej}$ equal to 89° and 90°. It was due to the fact that at $i_{ej} = 90°$ the vectors of initial velocities of the bodies considered were practically the same. In (Reyes-Ruiz et al., 2012) for $h = 100$ km values of $k_{pE/pM}$ were in the range from 30 to 33 at $v_{ej}$ equal to 12.7, 14.7, and 16.4 km/s, and were equal to 79 and 144 at $v_{ej}$ equal to 11.71 and 11.22 km/s. As noted in the previous section, the values of the ratio of the probabilities of collisions of bodies with the Earth and the Moon may be slightly less than the values of $k_{pE/pM}$.

In Figs. 1–3, the data are given for a final moment in time (usually equal to several hundred million years), when all bodies fell onto the planets or the Sun or reached 2000 AU from the Sun. Values of $k_{pE/pM}$ were generally larger when considering smaller time intervals. At $T = 10$ million years in approximately 90% of variants, the values of $k_{pE/pM}$ were more than at $T = T_{end}$, and the positive difference between these values did not exceed two in more than half of the variants, and was from two to four in about a quarter of the variants.

About the same (20–40) characteristic ratio of the number of bodies colliding with the Earth to the number of bodies colliding with the Moon was obtained in (Marov and Ipatov, 2023) for planetesimals that came to the Earth from the feeding zone of the terrestrial planets. For bodies initially located at a distance from the Sun greater than 3 AU, this ratio was mainly in the range from 16.4 to 17.4.

Figure 2 shows values of the probability of collisions of bodies with the Moon, calculated by the formula $p_M = p_E / k_{pE/pM}$, depending on the ejection angle $i_{ej}$ for a range of values of ejection velocity $v_{ej}$ and various ejection points. These probability values are given for the entire time interval considered. Probability $p_M$ of a collision of the body with the Moon was mainly in

the region of 0.01–0.017 at $v_{ej}$ = 11.3 km/s, 0.008–0.014 at $v_{ej}$ = 11.4 km/s, 0.008–0.014 at $v_{ej}$ = 11.5 km/s, 0.006–0.01 at $v_{ej}$ = 12 km/s and 0.004–0.008 at $v_{ej}$ ≥ 14 km/s. There was a noticeable dependence of $p_M$ on $i_{ej}$ only for the ejection point $W$ on the front hemisphere with $v_{ej}$ ≥ 14 km/s. For this ejection point the values of $p_M$ were noticeably less than the above values with $i_{ej}$ ≥ 60° and $v_{ej}$ = 14 km/s, with $i_{ej}$ ≥ 45° and $v_{ej}$ = 16.4 km/s, and also when $v_{ej}$ = 20 km/s. In (Reyes-Ruiz et al., 2012) for $h$ = 100 km and $v_{ej}$ = 12.7 km/s, a greater number of bodies collided with planets when bodies were ejected from the rear hemisphere of the Earth than from the front hemisphere. Due to the motion of the Earth in its orbit, bodies launched from its rear hemisphere have a lower velocity relative to the Sun, receive less eccentric heliocentric orbits and, accordingly, have a higher probability of collisions with planets.

As noted in the previous section, Fig. 2 shows the values of $p_M$ for the Moon's orbit at the boundary of the Earth's sphere of action. For the present orbit of the Moon (with $a_M$ = 60) these estimates for $p_M$ will be slightly larger (since the values of the ratio $k$ of the probabilities of collisions with the Earth and the Moon may be slightly less then $k_{pE/pM}$), but almost the same. In Table 1, with $a_M$ = 60 the ratio $k_{pE/pM}$ was in the ranges of 0.825–0.87, 0.71–0.78 and 0.65–0.74 with $k_{pE/pM}$ equal to 20, 30, and 40, respectively. The actual ratio of values of $k_{pE/pM}$ may be greater than these values, but does not exceed 1.

Above are the probabilities of collisions of ejected bodies with the Moon for the entire considered time interval (about several hundred million years). The difference between values of $k_{pE/pM}$ at $T$ = 10 million years and $T$ = $T_{end}$ generally did not exceed 10%. Therefore, for approximate estimates one can use the values of $k_{pE/pM}$ from Fig. 1. The fractions of bodies that fell on the Moon at different times relative to their final values are approximately the same as the data in Table 2 given for the Earth.

The ratio of the probabilities of collisions of bodies with the Moon in the first 10 million years to these probabilities for the entire time interval was mainly in the intervals of 0.7–0.8, 0.55–0.65 and 0.45–0.6 with $v_{ej}$ equal to 11.5, 12, and 16.4 km/s, respectively, (see Table 2). This ratio tended to decrease at higher values of $v_{ej}$.

Values of collision velocities of bodies with the Moon and Earth depending on the ejection angle $i_{ej}$ for a range of values $v_{ej}$ of the ejection velocities and various ejection points are shown in Fig. 3. The parabolic velocity on the surface of the Moon is several times less than the characteristic velocity of the body entering the sphere of action of the Earth. Therefore, the velocity at which the body collided with the Moon was only slightly higher than the velocity at which the body entered the sphere of action of the Earth. The ratio of these velocities is 1.05, 1.03 and 1.01 at $k_{pE/pM}$ equal to 20, 30, and 40, respectively.

The average velocity of collisions of ejected bodies with the Earth and the Moon is greater, the greater is the ejection velocity $v_{ej}$. The collision velocities of bodies with the Earth were mainly about 13, 14–15, 14–16, 14–20, 14–25 and 15–35 km/s with ejection velocities equal to 11.3, 11.5, 12, 14, 16.4 and 20 km/s, respectively. The collision velocities of bodies with the Moon were mainly in the range of 7–8, 10–12, 10–16, 11–23 and 12–33 km/s with $v_{ej}$ equal to 11.3, 12, 14, 16.4, and 20 km/s, respectively. The greater is $v_{ej}$, the wider is the range of average collision velocities with the Earth or Moon obtained for different angles and ejection points.

## GROWTH OF THE MOON EMBRYO WHEN MATERIAL EJECTED FROM THE EARTH'S EMBRYO FELL ON IT

Bodies ejected from the Earth could have participated in the formation of the Moon, both during the accumulation of the Earth and the Moon, and during the late heavy bombardment. In order to contain its current fraction of iron, the Moon must have accumulated the bulk of its mass from the Earth's mantle (Ipatov, 2018). The above calculations show that with the current orbit and mass of the Moon, about 1% of the bodies ejected from the Earth would fall on it. This fraction is smaller when the mass of the Moon embryo is smaller. From the estimates of the probabilities of collisions of bodies ejected from the Earth with the Moon moving in its present orbit (Fig. 2), we can conclude that in order for the Moon to acquire most of its mass from matter ejected from the Earth during its repeated bombardment, the mass of matter ejected from the Earth during its accumulation should be comparable to the mass of the Earth. Analyzing the data presented in the introduction on the characteristic velocities of collisions of planetesimals with the Earth and the dependence of the mass of matter ejected during a collision of a body with the Earth on the collision velocity, we can expect that in most collisions of impactors with the Earth the mass of the ejected matter was several times (usually an order of magnitude) less than the mass of the impactor. An ejection of matter comparable to the mass of the impactor could only occur during relatively rare collisions with the Earth of bodies coming from beyond the orbit of Mars. These estimates suggest that the bodies ejected from the Earth and falling onto the growing Moon were probably not enough to grow the Moon from a small embryo in its present orbit. This result argues in favor of the formation of a lunar embryo and its further growth to most of the present mass of the Moon near the Earth. Ipatov (2018) suggested that the initial embryo of the Moon with a mass of no more than 0.1 of the mass of the Moon was formed simulta-neously with the embryo of the Earth from a common rarefied condensation.

The ratio of the fraction of bodies colliding with the Earth to the fraction of bodies colliding with the Moon for planetesimals from the feeding zones of terrestrial planets is close to the similar ratio $k_{pE/pM}$ considered above for bodies ejected from the Earth and leaving its sphere of action. In (Marov and Ipatov, 2021) it was noted that the ratio $k_{E.M.}$ of the number of planetesimals colliding with the Earth and Moon varied mainly from 20 to 40. Moreover, for planetesimals with semimajor axes of initial orbits of $0.9 \le a_o \le 1.1$ AU and with small initial eccentricities, $k_{E.M.}$ was mainly in the range from 30 to 40, and planetesimals, initially more distant from the Earth's orbit, came to it from more eccentric orbits, and this ratio $k_{E.M.}$ for them was less than for close, weakly eccentric orbits. In Fig. 1, values of $k_{pE/pM}$ ranged mainly from 35–47 with $v_{ej}$ = 11.3 km/s to 17–30 with $v_{ej}$ = 16.4 km/s and 15–25 with $v_{ej}$ = 20 km/s.

The total mass of bodies ejected from the Earth and colliding with the Moon in its present orbit before leaving its sphere of action was probably small. In the calculations carried out, with the exception of the ejection of bodies with velocities slightly exceeding the parabolic velocity of the Earth, the ejected bodies immediately left the Earth's sphere of action and could cross the Moon's orbit no more than once before leaving this sphere. Many inclined orbits of ejected bodies may not have crossed the Moon's orbit. The ratio of the two radii of the Moon to the length of its orbit is 0.00144 and indicates the maximum value of the probability of a collision of a body with the Moon for the model under consideration. This ratio is an order of magnitude smaller than those shown in Fig. 2 of probabilities of collisions of ejected bodies that left the Earth's sphere of action, but then repeatedly entered this sphere over the course of millions of years. For the possibility of multiple approaches of an ejected body with the Moon, it is necessary that the apogee of its orbit around the Earth be beyond the orbit of the Moon, but within the sphere of action of the Earth (and the pericentric distance should be greater than the radius of the Earth). This possibility is greater for the smaller distance of the Moon from the Earth.

Let us consider the following model. Let the average mass of ejected matter during collisions of planetesimals with the Earth be equal to $k_{ej}$ of the mass of matter entering the Earth. As the Earth's mass increases by $k_E \times m_E$ the total mass of planetesimals that entered the sphere of action of the Earth and collided with the Moon is equal to $k_E m_E (1 + k_{ej})/k_{E.M.}$. The mass of bodies ejected from the Earth and falling onto the Moon after leaving the Earth's sphere of action (with possible multiple subsequent hits within the Earth's sphere of action) is equal to $k_{ej} \times k_E \times m_E \times p_E/k_{pE/pM}$. The ratio $k_{plb}$ of the mass of planetesimals that fell directly on the Moon to the mass of bodies ejected from the Earth and that fell on the Moon is equal to $(k_{pE/pM}/k_{E.M.})(1 + k_{ej})/(k_{ej} \times p_E)$. As noted above, $k_{pE/pM}/k_{E.M.}$ is close to 1. That is why $k_{plb}$ is close to $(1 + k_{ej})/(k_{ej} \times p_E)$. Because $k_{ej} < 1$ (otherwise the Earth would not have grown), then $(1 + k_{ej})/k_{ej} > 2$. The fraction $p_E$ of bodies ejected from the Earth and later falling back onto it, is about 0.2–0.3 (Tables 2, 3). From the results presented in the Introduction, we can conclude that for most collisions of planetesimals with the Earth $k_{ej}$ did not exceed 0.15. With $k_{ej}$ = 0.15 and $p_E$ = 0.3, the value of $k_{plb}$ is close to 25. These estimates indicate that the total mass of planetesimals (with more iron than in the Earth's mantle) that collided with the Moon in its present orbit was at least several times greater than the mass of the bodies ejected from the Earth and then collided with the Moon. Taking into account the fall of planetesimals richer in iron directly onto the Moon, as well as the estimates at the beginning of this section, testifies in favor of the Moon acquiring the bulk of the material closer to the Earth than the present orbit of the Moon.

With the Moon's orbit close to the Earth, the fraction of bodies ejected from the Earth and leaving the Earth's sphere of action when falling on the Moon could be only slightly more than 13 (the ratio of the squares of the radii of the Earth and the Moon). However, the model of the growth of the mass of the Moon embryo mainly due to such bodies is hampered by the fact that an even larger number of bodies (not ejected from the Earth) with approximately the same (on average slightly higher) velocities could have fallen directly onto the Moon (without being ejected from the Earth) and bring to the Moon material richer in iron than the Earth's mantle.

It seems more likely to us that, as was proposed in (Ipatov, 2018), the initial embryo of the Moon (with a mass an order of magnitude smaller than the present mass of the Moon) was formed simultaneously with the embryo of the Earth during the compression of a general rarefied condensation. This condensation received the angular momentum necessary for the formation of a binary system when two condensations collided. Such collisions of condensations are not frequent, which explains the absence of large satellites of other terrestrial planets. The growth of the Moon embryo probably occurred within the framework of the multi-impact model, when matter ejected from the Earth formed a disk near the Earth from which bodies accumulated. In (Rufu et al., 2017), the masses of the impactors were no less than the mass of the present Moon. Our calculations of the evolution of the orbits of ejected bodies were for collisions of smaller planetesimals with the Earth, in which most of the ejected matter that did not immediately fall back to the Earth left the Earth's sphere of action. The presence of an initially large lunar embryo contributed to a more efficient capture of the material ejected from the Earth by the lunar embryo. For more efficient growth of the Moon embryo, it is desirable that during some collisions of impactor bodies with the Earth, the ejected bodies would not simply fly out of the crater, but some of the matter would enter in an orbit around the Earth.

## CONCLUSIONS

The evolution of the orbits of bodies ejected from the Earth after impacts of large planetesimals (bodies) was considered. Such collisions occurred during the Earth's accumulation and during the Late Heavy Bombardment. In each variant of the calculations, the motion of 250 bodies ejected from the Earth was studied at fixed values of the ejection angle $i_{ej}$ (measured from the surface plane), ejection velocity $v_{ej}$ and the time integration step. In different variants, the values of the ejection angle $i_{ej}$ were 15°, 30°, 45°, 60°, 89° or 90°. The velocity $v_{ej}$ of bodies ejected from the Earth were generally equal to 11.22, 11.5, 12, 14, 16.4 or 20 km/s, but other values in the range from 11.22 to 11.5 km/s were also considered. The gravitational influence of the Sun and all eight planets was taken into account. Bodies that collided with planets or the Sun or reached 2000 AU from the Sun were excluded from the integration. The ejection of bodies from six opposite points on the Earth's surface was considered for a number of values of velocities and angles of ejection of bodies. In the considered variants for calculating the motion of bodies ejected from the Earth, most of the bodies left the Hill sphere of the Earth and moved in heliocentric orbits. Their dynamical lifetime reached several hundred million years.

The average velocity of collisions of ejected bodies with the Earth and the Moon is greater, the greater is the ejection velocity. The collision velocities of bodies with the Earth were about 13, 14–15, 14–16, 14–20, 14–25, and 15–35 km/s with ejection velocities equal to 11.3, 11.5, 12, 14, 16.4, and 20 km/s, respectively. The collision velocities of bodies with the Moon were mainly in the range of 7–8, 10–12, 10–16, 11–23, and 12–33 km/s with $v_{ej}$, equal to 11.3, 12, 14, 16.4, and 20 km/s, respectively.

Probabilities $p_E$ and $p_M$ of collisions of bodies with the Earth and Moon were on average lower at higher ejection velocities $v_{ej}$. Over the entire time interval under consideration with values of $v_{ej}$ equal to 11.5, 12, and 14 km/s, values of $p_E$ were approximately 0.3, 0.2, and 0.15–0.2, respectively. With ejection velocities $v_{ej} \leq 11.25$ km/s, i.e., slightly exceeding the parabolic velocity, most of the ejected bodies fell back to the Earth. The probability $p_M$ of the collision of a body ejected from the Earth with the Moon was approximately 15–35 times less than with the Earth with the ejection velocity $v_{ej} \geq 11.5$ km/s. The probability of collisions of such bodies with the Moon in its present orbit was mainly about 0.004–0.008 with ejection velocities of at least 14 km/s and about 0.006–0.01 with $v_{ej}$ = 12 km/s. This probability was greater at lower ejection velocities and was in the range of 0.01–0.02 with ejection velocities of 11.3 km/s.

The results obtained show that the bodies ejected from the Earth and falling on the embryo of the Moon would not have been enough for the Moon to grow to its present mass from a small embryo moving along the present orbit of the Moon. This conclusion indicates that the Moon's embryo was formed (for example, simultaneously with the Earth's embryo from a general rarefied condensation) and subsequently acquired most of the Moon's mass at a short distance from the Earth. For more efficient growth of the Moon embryo, it is desirable that during some collisions of impactor bodies with the Earth, the ejected bodies do not simply fly out of the crater, but some of the matter enters orbit around the Earth, as in the multi-impact model.

## ACKNOWLEDGMENTS

The author expresses deep gratitude to A.B. Makalkin for useful comments that helped to improve the paper.

## FUNDING

The research was supported by the Russian Science Foundation, project 21-17-00120.

## CONFLICT OF INTEREST

The authors of this work declare that they have no conflicts of interest.

## REFERENCES

Afanas'ev, V.N. and Pechernikova, G.V., On the probability of capturing preplanetary bodies into the protolunar swarm during the formation of the Earth–Moon system, *Sol. Syst. Res.,* 2022, vol. 56, no. 6, pp. 382–402.

Armstrong, J.C., Wells, L.E., and Guillermo, G., Rummaging through Earth's attic for remains of ancient life, *Icarus,* 2002, vol. 160, pp. 183–196.

Artemieva, N., From the Moon to the Earth without Jules Verne—lunar meteorites and lunar dust delivery, *45th Lunar and Planet. Sci. Conf.,* 2014, p. 1659.

Artemieva, N.A. and Shuvalov, V.V., Numerical simulation of high-velocity impact ejecta following falls of comets and asteroids onto the Moon, *Sol. Syst. Res.,* 2008, vol. 42, no. 4, pp. 329–334.

Barnouin, O.S., Dalya, R.T., Cintala, M.J., and Crawford, D.A., Impacts into coarse-grained spheres at moderate impact velocities: implications for cratering on asteroids and planets, *Icarus,* 2019, vol. 325, pp. 67–83.

Barr, A.C., On the origin of Earth's Moon, *J. Geophys. Res.: Planets,* 2016, vol. 121, pp. 1573–1601. https://arxiv.org/pdf/1608.08959.pdf.

Beech, M., Comte, M., and Coulson, I., The production of terrestrial meteorites—Moon accretion and lithopan-spermia, *Am. J. Astron. Astrophys.,* 2019, vol. 7, no. 1, pp. 1–9.

Bottke, W.F. and Norman, M.D., The late heavy bombardment, *Ann. Rev. Earth Planet. Sci.,* 2017, vol. 45, pp. 619–647.

Bulirsh, R. and Stoer, J., Numerical treatment of ordinary differential equations by extrapolation methods, *Num. Math.,* 1966, no. 8, pp. 1–13.

Cameron, A.G.W. and Ward, W.R., The origin of the Moon, *Lunar and Planet. Sci. Conf.,* 1976, vol. 7, pp. 120–122.

Canup, R.M., Simulations of a late lunar-forming impact, *Icarus,* 2004, vol. 168, no. 2, pp. 433–456.

Canup, R.M., Forming a Moon with an Earth-like composition via a giant impact, *Science,* 2012, vol. 338, pp. 1052–1055.

Canup, R.M. and Asphaug, E., Origin of the Moon in a giant impact near the end of the Earth's formation, *Nature,* 2001, vol. 412, no. 6848, pp. 708–712.

Canup, R.M., Barr, A.C., and Crawford, D.A., Lunar-forming impacts: High-resolution SPH and AMR-CTH simulations, *Icarus,* 2013, vol. 222, pp. 200–219.

Canup, R.M., Righter, K., Dauphas, N., Pahlevan, K., Ćuk, M., Lock, S.J., Stewart, S.T., Salmon, J., Rufu, R., Nakajima, M., and Magna, T., Origin of the Moon, in *New Views of the Moon II,* 2021, p. 71. https://arxiv.org/abs/2103.02045.

Chambers, J.E., A hybrid symplectic integrator that permits close encounters between massive bodies, *Mon. Not. R. Astron. Soc.,* 1999, vol. 304, pp. 793–799.

Citron, R.I., Aharonson, O., Perets, H., and Genda, H., Moon formation from multiple large impacts, *45th Lunar and Planet. Sci. Conf.,* 2014, p. 2085.

Cuk, M. and Stewart, S.T., Making the Moon from a fast-spinning Earth: A giant impact followed by resonant despinning, *Science,* 2012, vol. 338, pp. 1047–1052.

Cuk, M., Hamilton, D.P., Lock, S.J., and Stewart, S.T., Tidal evolution of the Moon from a high-obliquity, high-angular-momentum Earth, *Nature,* 2016, vol. 539, pp. 402–406.

Cuzzi, J.N. and Hogan, R.C., Primary accretion by turbulent concentration: The rate of planetesimal formation and the role of vortex tubes, *43rd Lunar and Planet. Sci. Conf.,* 2012, p. 2536.

Cuzzi, J.N., Hogan, R.C., and Sharif, K., Toward planetesimals: Dense chondrule clumps in the protoplanetary nebula, *Astrophys. J.,* 2008, vol. 687, pp. 1432–1447.

Cuzzi, J.N., Hogan, R.C., and Bottke, W.F., Towards initial mass functions for asteroids and kuiper belt objects, *Icarus,* 2010, vol. 208, pp. 518–538.

Frantseva, K., Nesvorny, D., Mueller, M., van der Tak, F.F.S., Kate, I.L., and Pokorny, P., Exogenous delivery of water to Mercury, *Icarus,* 2022, vol. 383, p. 114980.

Galimov, E.M., The problem of the origin of the Earth–Moon system, in *Problemy zarozhdeniya i evolyutsii biosfery* (Problems of the Origin and Evolution of the Biosphere), Galimov, E.M., Ed., Moscow: Nauka, 1995, pp. 8–45.

Galimov, E.M., Current state of the problem of the origin of the Earth–Moon system, in *Problemy zarozhdeniya i evolyutsii biosfery* (Problems of the Origin and Evolution of the Biosphere), Galimov, E.M., Ed., Moscow: LIBROKOM, 2008, pp. 213–222.

Galimov, E.M., Formation of the Moon and the Earth from a common supraplanetary gas-dust cloud, *Geochem. Int.,* 2011, vol. 49, pp. 537–554.

Galimov, E.M., Analysis of isotope systems (Hf-W, Rb-Sr, J-Pu-Xe, U-Pb) in relation to the problem of planet formation using the example of the Earth–Moon system, in *Problemy zarozhdeniya i evolyutsii biosfery* (Problems of the Origin and Evolution of the Biosphere), Galimov, E.M., Ed., Moscow: KRASAND, 2013, pp. 47–59.

Galimov, E.M. and Krivtsov, A.M., *Origin of the Moon. New Concept,* Berlin: De Gruyter, 2012, 168 p.

Galimov, E.M., Krivtsov, A.M., Zabrodin, A.V., Legkostupov, M.S., Eneev, T.M., and Sidorov, Yu.I., Dynamic model for the formation of the Earth–Moon system, *Geochem. Int.,* 2005, vol. 43, no. 11, pp. 1045–1055.

Gallant, J., Gladman, B., and Cuk, M., Current bombardment of the Earth-Moon system: Emphasis on cratering asymmetries, *Icarus,* 2009, vol. 202, pp. 371–382.

Gattacceca, J., Debaille, V., Devouard, B., Leya, I., Jourdan, F., Braucher, R., Roland, J., Pourkhorsandi, H., Goderis, S., and Jambon, A., Northwest Africa 13188: A possible meteorite ... from Earth!, *Abstracts of Goldschmidt Conf.,* 2023. https://conf.goldschmidt.info/goldschmidt/2023/meetingapp.cgi/Paper/20218.

Gladman, B., Dones, L., Levison, H.F., and Burns, J.A., Impact seeding and reseeding in the inner Solar System, *Astrobiology,* 2005, vol. 5, pp. 483–496. https://www.liebertpub.com/doi/10.1089/ast.2005.5.483.

Gor'kavyi, N.N., Formation of the Moon and double asteroids, *Izv. Krym. Astrofiz. Obs.,* 2007, vol. 103, no. 2, pp. 143–155.

Gorkavyi, N.N., The new model of the origin of the Moon, *Bull. Am. Astron. Soc.,* 2004, vol. 36, p. 861.

Hartmann, W.K. and Davis, D.R., Satellite-sized planetesimals and lunar origin, *Icarus,* 1975, vol. 24, pp. 504–515.

Holsapple, K.A. and Housen, K.R., A crater and its ejecta: An interpretation of deep impact, *Icarus,* 2007, vol. 187, pp. 345–356.

Ipatov, S.I., Evolution times for disks of planetesimals, Sov. Astron., 1988, vol. 32 (65), no. 5, pp. 560–566. https://www.academia.edu/44452977/Evolution_times_for_disks_of_ planetesimals.

Ipatov, S.I., *Migratsiya nebesnykh tel v Solnechnoi sisteme* (Migration of Celestial Bodies in the Solar System), Moscow: URSS, 2000.
https://elibrary.ru/item.asp?id=46237738.
http://www.rfbr.ru/rffi/ru/books/o_29239
https://doi.org/10.17513/np.451.

Ipatov, S.I., Formation of trans-Neptunian satellite systems at the stage of condensations, *Sol. Syst. Res.,* 2017a, vol. 51, no. 4, pp. 294–314. https://arxiv.org/abs/1801.05217. https://doi.org/10.1134/S0038094617040013

Ipatov, S.I., Origin of orbits of secondaries in the discovered trans-Neptunian binaries, *Sol. Syst. Res.,* 2017b, vol. 51, no. 5, pp. 409–416. https://arxiv.org/abs/1801.05254. https://doi.org/10.1134/S0038094617050045

Ipatov, S.I., Formation of embryos of the Earth and the Moon from the common rarefied condensation and the subsequent growth, *Sol. Syst. Res.,* 2018, vol. 52, no. 5, pp. 401–416. http://arxiv.org/abs/2003.09925. https://www.academia.edu/38545653/. https://doi.org/10.1134/S0038094618050040

Ipatov, S.I., Probabilities of collisions of planetesimals from different regions of the feeding zone of the terrestrial planets with forming planets and the Moon, *Sol. Syst. Res.,* 2019, vol. 53, no. 5, pp. 332–361. http://arxiv.org/abs/2003.11301. https://rdcu.be/bRVA8. https://www.academia.edu/42216860. https://doi.org/10.1134/S0038094619050046

Ipatov, S.I., Collisions of bodies ejected from several places on the Earth and the Moon with the terrestrial planets and the Moon, *54nd Lunar and Planet. Sci. Conf.,* 2023, p. 1508. https://www.hou.usra.edu/meetings/lpsc2023/pdf/1508.pdf.

Ipatov, S.I. and Mather, J.C., Migration of trans-Neptunian objects to the terrestrial planets, *Earth, Moon, and Planets,* 2003, vol. 92, pp. 89–98. http://arXiv.org/format/astro-ph/0305519. https://doi.org/10.1023/B:MOON.0000031928.45965.7b

Ipatov, S.I. and Mather, J.C., Comet and asteroid hazard to the terrestrial planets, *Adv. Space Res.,* 2004a, vol. 33, pp. 1524–1533. http://arXiv.org/format/astro-ph/0212177. https://doi.org/10.1016/S0273-1177(03)00451-4

Ipatov, S.I. and Mather, J.C., Migration of Jupiter-family comets and resonant asteroids to near-Earth space, *Ann. NY Acad. Sci.,* 2004b, vol. 1017, pp. 46–65. http://arXiv.org/format/astro-ph/0308448. https://doi.org/10.1196/annals.1311.004

Johansen, A., Oishi, J.S., Mac Low, M.-M., Klahr, H., Henning, T., and Youdin, A., Rapid planetesimal formation in turbulent circumstellar disks, *Nature,* 2007, vol. 448, pp. 1022–1025.

Johansen, A., Youdin, A., and Klahr, H., Zonal flows and long-lived axisymmetric pressure bumps in magnetorotational turbulence, *Astrophys. J.,* 2009a, vol. 697, pp. 1269–1289.

Johansen, A., Youdin, A., and Mac Low, M.-M., Particle clumping and planetesimal formation depend strongly on metallicity, *Astrophys. J.,* 2009b, vol. 704, pp. L75–L79.

Johansen, A., Klahr, H., and Henning, T., High-resolution simulations of planetary formation in turbulent protoplanetary discs, *Astron. Astrophys.,* 2011, vol. 529, p. A62.

Johansen, A., Youdin, A.N., and Lithwick, Y., Adding particle collisions to the formation of asteroids and kuiper belt objects via streaming instabilities, *Astron. Astrophys.,* 2012, vol. 537, p. A125.

Levison, H.F. and Duncan, M.J., The long-term dynamical behavior of short-period comets, *Icarus,* 1994, vol. 108, pp. 18–36.

Lyra, W., Johansen, A., Klahr, H., and Piskunov, N., Embryos grown in the dead zone. assembling the first protoplanetary cores in low mass self-gravitating circumstellar disks of gas and solids, *Astron. Astrophys.,* 2008, vol. 491, pp. L41–L44.

Lyra, W., Johansen, A., Zsom, A., Klahr, H., and Piskunov, N., Planet formation bursts at the borders of the dead zone in 2d numerical simulations of circumstellar disks, *Astron. Astrophys.,* 2009, vol. 497, pp. 869–888.

Marov, M.Ya. and Ipatov, S.I., Formation of the Earth and Moon: Influence of small bodies, *Geochem. Int.,* 2021, vol. 59, no. 11, pp. 1010–1017. https://doi.org/10.1134/S0016702921110070

Marov, M.Ya. and Ipatov, S.I., Migration processes in the Solar System and their role in the evolution of the Earth and planets, *Phys. Usp.,* 2023, vol. 66, no. 1, pp. 2–31. https://doi.org/10.3367/UFNe.2021.08.039044

Marov, M.Ya., Voropaev, S.A., Ipatov, S.I., Badyukov, D.D., Slyuta, E.N., Stennikov, A.V., Fedulov, V.S., Dushenko, N.V., Sorokin, E.M., and Kronrod, E.V., *Formirovanie Luny i rannyaya evolyutsiya Zemli* (Formation of the Moon and Early Evolution of the Earth), Moscow: URSS, 2019, 320 p.

Melosh, H., *Impact Cratering: A Geologic Process,* New York: Oxford Univ. Press, 1989.

Melosh, H., Ejection of rock fragments from planetary bodies, *Geology,* 1985, vol. 13, pp. 144–148.

Őpik, E.J., Collision probabilities with the planets and the distribution of interplanetary matter, *Proc. R. Irish Acad. Sect. A,* 1951, vol. 54, pp. 165–199.

Pahlevan, K. and Morbidelli, A., Collisionless encounters and the origin of the lunar inclination, *Nature,* 2015, vol. 527, no. 7579, pp. 492–494. https://arxiv.org/abs/1603.06515.

Raducan, S.D., Davison, T.M., and Collins, G.S., Ejecta distribution and momentum transfer from oblique impacts on asteroid surfaces, *Icarus,* 2022, vol. 374, p. 114793. https://arxiv.org/abs/2105.01474.

Reyes-Ruiz, M., Chavez, C.E., Aceves, H., Hernandez, M.S., Vazquez, R., and Nuñez, P.G., Dynamics of escaping Earth ejecta and their collision probabilities with different solar system bodies, *Icarus,* 2012, vol. 220, pp. 777–786.

Ringwood, A.E., Flaws in the giant impact hypothesis of lunar origin, *Earth Planet. Sci. Lett.,* 1989, vol. 95, nos. 3–4, pp. 208–214.

Rufu, R. and Aharonson, O., A multiple impact hypothesis for Moon formation, *46th Lunar and Planet. Sci. Conf.,* 2015, p. 1151.

Rufu, R., Aharonson, O., and Perets, H.B., A multiple-impact origin for the Moon, *Nat. Geosci.,* 2017, vol. 10, pp. 89–94.

Rufu, R., Salmon, J., Pahlevan, K., Visscher, C., Nakajima, M., and Righter, K., The origin of the Earth-Moon system as revealed by the Moon, *Bull. Am. Astron. Soc.,* 2021, vol. 53, no. 4. https://doi.org/10.3847/25c2cfeb.6e7d4ab6

Ruskol, E.L., The origin of the Moon. I. Formation of a swarm of bodies around the Earth, *Sov. Astron.,* 1961, vol. 4, no. 4, p. 657.

Ruskol, E.L., On the origin of the Moon. II. The growth of the Moon in the circumterrestrial swarm of satellites, *Sov. Astron.,* 1963, vol. 7, no. 2, p. 221.

Ruskol, E.L., The origin of the Moon. III. Some aspects of the dynamics of the circumterrestrial swarm, *Sov. Astron.,* 1971, vol. 15, no. 4, p. 646.

Ruskol, E.L., *Proiskhozhdenie Luny* (The Origin of the Moon), Moscow: Nauka, 1975, 188 p.

Salmon, J. and Canup, R.M., Lunar accretion from a Roche-interior fluid disk, *Astrophys. J.,* 2012, vol. 760, p. A83.

Shuvalov, V.V. and Trubetskaya, I.A., Numerical simulation of high-velocity ejecta following comet and asteroid impacts: preliminary results, *Sol. Syst. Res.,* 2011, vol. 45, pp. 392–401.

Svetsov, V., Cratering erosion of planetary embryos, *Icarus,* 2011, vol. 214, pp. 316–326.

Svetsov, V.V., Pechernikova, G.V., and Vityazev, A.V., A model of Moon formation from ejecta of macroimpacts on the Earth, *43th Lunar and Planet. Sci. Conf.,* 2012, p. 1808.

Touma, J. and Wisdom, J., Evolution of the Earth–Moon system, *Astron. J.,* 1994, vol. 108, no. 5, pp. 1943–1961.

Vasil'ev, S.V., Krivtsov, A.M., and Galimov, E.M., Study of the planet-satellite system growth process as a result of the accumulation of dust cloud material, *Sol. Syst. Res.,* 2011, vol. 45, no. 5, pp. 410–419.

Vityazev, A.V. and Pechernikova, G.V., Early differentiation of the Earth and the problem of lunar composition, *Fiz. Zemli,* 1996, no. 6, pp. 3–16.

Wang, N. and Zhou, J-L., Analytical formulation of lunar cratering asymmetries, *Astron. Astrophys.,* 2016, vol. 594, p. A52.

Youdin, A.N., On the formation of planetesimals via secular gravitational instabilities with turbulent stirring, *Astrophys. J.,* 2011, vol. 731, p. A99.

Youdin, A.N. and Kenyon, S.J., From disks to planets, in *Planets, Stars and Stellar Systems. Solar and Stellar Planetary Systems,* Oswalt, T.D., French, L.M., and Kalas, P., Eds., Dordrecht: Springer Science+Business Media, 2013, vol. 3, pp. 1–62. https://doi.org/10.1007/978-94-007-5606-9_1

Zharkov, V.N., On the history of the lunar orbit, *Sol. Syst. Res.,* 2000, vol. 34, pp. 1–11.